\documentclass[12pt,letterpaper]{article}

\usepackage{graphicx,array}
\usepackage{url}
\usepackage{color}
\usepackage{latexsym}
\usepackage{amsthm}
\usepackage{amsmath}
\usepackage{amssymb}
\usepackage{amsfonts}
\usepackage[numbers,sort&compress]{natbib}
\usepackage{bm}
\usepackage{slashed}
\usepackage{mathrsfs}
\usepackage{enumerate}
\usepackage{tikz}
\usepackage{siunitx}
\usepackage{mdframed}
\usepackage{setspace}  
\usepackage{esvect}
\usepackage{esint}

\usepackage[utf8x]{inputenc}%
\usepackage{tcolorbox}%

\usepackage{hyperref} 

\hypersetup{
    colorlinks=true,       
    linkcolor=red,          
    citecolor=blue,        
    filecolor=magenta,      
    urlcolor=blue           
}
\usepackage[all]{hypcap} 

\usepackage{natbib}
\newcommand{\nc}{\newcommand}

\nc{\beq}{\begin{equation}}  
\nc{\eeq}{\end{equation}}  
\nc{\beqa}{\begin{eqnarray}}  
\nc{\eeqa}{\end{eqnarray}}  
\nc{\bit}{\begin{itemize}}  
\nc{\eit}{\end{itemize}}  

\usepackage{floatrow}
\newfloatcommand{capbtabbox}{table}[][\FBwidth]

\usepackage{blindtext}

\title{ 
{\bf Hairy Magnetic and Dyonic Black Holes 
\\ in the Standard Model
}
\author{\large Yang Bai and Mrunal Korwar}
\date{\small \it 
Department of Physics, University of Wisconsin-Madison, Madison, WI 53706, USA
}
}

\begin{document}

\maketitle

\setlength{\parskip}{0.2ex}

\begin{abstract}	
Spherically symmetric magnetic and dyonic black holes with a magnetic charge $Q=2$ are studied in the Standard Model and general relativity. A magnetically charged black hole with mass below $9.3\times 10^{35}$~GeV has a ``hairy" cloud of electroweak gauge and Higgs fields outside the event horizon with $1/m_W$ in size. An extremal magnetic black hole has a hair mass of 3.6 TeV,  while an extremal dyonic black hole has an additional mass of $q^2 \times 1.6$~GeV for a small electric charge $q \ll 2\pi/e^2$. A hairy dyonic black hole with an integer charge is not  stable and can decay into a magnetic one plus charged fermions. On the other hand, a hairy magnetic black hole can evolve via Hawking radiation into a nearly extremal one that is cosmologically stable and an interesting object to be searched for. 
\end{abstract}

\thispagestyle{empty}  
\newpage  
  
\setcounter{page}{1}  



\section{Introduction}
\label{sec:intro}
Although the black hole no-hair conjecture has been proposed a long time ago~\cite{Misner:1974qy}, various violations have been found in the systems like Einstein-Skyrme~\cite{Droz:1991cx,Heusler:1991xx,Bizon:1992gb,Dvali:2016mur} and 
Einstein-Yang-Mills-Higgs~\cite{Lee:1991qs,Lee:1991vy,Greene:1992fw,Maeda:1993ap,Lee:1994sk,Tachizawa:1994wn,Weinberg:1994eb} (see \cite{Cardoso:2016ryw} for a review). For instance, Refs.~\cite{Lee:1991qs,Lee:1991vy} have demonstrated that the magnetically charged Reissner-Nordstr\"om (RN) black holes is classically unstable and can evolve into a hairy magnetic black hole with a cloud of gauge and Higgs field configures. The existence of a non-trivial hair for the black hole changes the black hole properties in various ways including the faster and effectively two-dimensional Hawking radiation~\cite{Maldacena:2020skw}. Given the discovery of the Higgs boson in 2012~\cite{Chatrchyan:2012ufa,Aad:2012tfa} and the completeness of the Standard Model (SM), we study the properties of magnetic and dyonic black holes with electroweak gauge and Higgs hair. In our study, we will consider the magnetic charge $Q=2$ case with a spherically symmetric configuration for the whole system. We want to emphasize that the existence of hairy magnetic and dyonic black holes only depends on the known SM and general relativity, although their abundance in the current Universe is subject to the early-universe formation mechanism and various phenomenological constraints~\cite{Bai:2020spd}. 

In the direction of searching for localized and magnetically charged object with finite energy, there is no viable candidate just based on the SM interactions (see \cite{Cho:1996qd,Ellis:2016glu} for studies including physics beyond the SM). The existence of a black hole inside a magnetic monopole can make the energy of the whole system finite, because the divergent behavior of the monopole energy in the small radius region is regularized by the black hole event horizon. So, the hairy magnetic black hole (hMBH) serves as a unique and justified magnetically charged object based on experimentally established SM (on the purely theoretically side, we may have the t 'hooft-Polyakov monopole~\cite{tHooft:1974kcl,Polyakov:1974ek} in the grand unified theories~\cite{Georgi:1974sy}). 

This paper is organized as follows. We first derive the equations of motion for a general hairy dyonic black hole (hDBH) in Section~\ref{sec:lagrangian}. In Section~\ref{sec:magnetic-mass-profile}, we solve the equations of motion with proper boundary conditions (BCs) and calculate the mass and the field profiles for the hairy magnetic black hole case. We discuss the Hawking temperature and evolution of non-extremal hairy magnetic black hole in Section~\ref{sec:temperature-evolution}. In Section~\ref{sec:dyonic-mass-profile}, the mass and field profiles for the hairy dyonic black holes will be calculated with some of their properties discussed in Section~\ref{sec:dyon-properties}. We conclude our paper and discuss some future directions in Section~\ref{sec:conclusion}. 

\section{Lagrangian and equations of motion}
\label{sec:lagrangian}
Starting with the Einstein-Hilbert action in general relativity and including the matter Lagrangian in the SM of particle physics, one has the action
\beqa
S = \int d^4 x \sqrt{-g} \left[ - \frac{1}{16\pi\,G} \, R + \mathcal{L}_{\rm SM} \right] ~. 
\eeqa
Here, $G = 1/M^2_{\rm pl}$ with $M_{\rm pl} = 1.22 \times 10^{19}$~GeV is the Newton constant; $R$ is the Ricci scalar; $g\equiv \mbox{det}(g_{\mu\nu})$ is the determinant of the metric tensor matrix; the metric $g_{\mu\nu} = (g^{\mu\nu})^{-1}$ has the convention of $(+1, -1, -1, -1)$ for a flat-space metric. For the matter Lagrangian, we only include the known SM Lagrangian with focus on the electroweak $SU(2)_W \times U(1)_Y$ gauge sector
\beqa
\mathcal{L}_{\rm SM} \supset \mathcal{L}_{\rm EW} =  - \frac{1}{4}\,W^a_{\mu\nu}W^{a\,\mu\nu}  - \frac{1}{4}Y_{\mu\nu}Y^{\mu\nu}  +  |D_\mu H|^2 - \frac{\lambda}{2}\left( H^\dagger H - \frac{v^2}{2} \right)^2 ~,
\eeqa
with $v = 246$~GeV and $\lambda\approx 0.26$ to have the Higgs boson mass of $m_h =\sqrt{\lambda}\,v \approx 125$~GeV~\cite{Aad:2015zhl}. Here, $W^a_\mu$ with $a=1,2,3$ are $SU(2)_W$ gauge bosons and $Y_\mu$ is the hypercharge gauge boson. The gauge field tensors are $W^a_{\mu\nu}= \partial_\mu W^a_\nu - \partial_\nu W^a_\mu - g\,\epsilon_{abc} W^b_\mu W^c_\nu$ and $Y_{\mu\nu}= \partial_\mu Y_\nu - \partial_\nu Y_\mu$. The covariant derivative of the Higgs doublet is 
\beqa
D_\mu H = \left(\partial_\mu - i\, \frac{g}{2}\,\sigma^a\,W^a_\mu - i\, \frac{g_{_Y}}{2} \,Y_\mu\right) H ~,
\eeqa
with $\sigma^a$ as the Pauli matrices and the two gauge couplings $g = e/\sin{\theta_W}$ and $g_{_Y} = e/ \cos{\theta_W}$ with $e = \sqrt{4\pi\,\alpha}$ and $\alpha \approx 1/128$ at the electroweak scale. Here, $\theta_W$ is the weak mixing angle with $\sin{\theta_W} \approx \sqrt{0.23}$. The constant term for the Higgs potential is chosen to have a zero value when the Higgs field sits at the potential minimum  $\langle H\rangle = (0, v/\sqrt{2})^T$. 

For both magnetic and dyonic black holes, we will consider only the $Q=2$ magnetic charge in this paper (we will use $Q$ to label the magnetic charge and $q$ for the electric charge). At a long distance, the magnetic field is $\mathbf{B}(\mathbf{r}) = Q\,e_{\rm M}\,\hat{\mathbf{r}}/(4\pi r^2)$ with the magnetic coupling $e_{\rm M} = 2\pi/e$ following the Dirac quantization for the minimum charge $Q=1$. For $Q=2$, we anticipate a spherically symmetric solution for both magnetic and dyonic black holes. Therefore, we parametrize the metric as
\beqa
\label{eq:metric}
ds^2 =  P^2(r)\,N(r) \, dt^2 - N(r)^{-1}\, dr^2 - r^2 \, d\theta^2 - r^2 \,\sin^2{\theta} \, d\phi^2 ~,
\eeqa
in the spherical coordinate. For the Einstein-Hilbert action and using integrating by parts that does not change the later equations of motion, one has (see \cite{VanNieuwenhuizen:1975tc} for a different metric convention) 
\beqa
S_{\rm E} = -\frac{1}{16\pi\,G} \int d^4 x \sqrt{-g} \,R = -\, \frac{1}{2\,G}  \int dt \,dr \, r\, P'(1 - N) ~,
\eeqa
with the prime denoting differentiation with respect to $r$. 

For the matter part and following Ref.~\cite{Cho:1996qd}, we use the following ansatz for a spherical monopole (or dyon) configuration in the hedgehog gauge~\footnote{The topological argument for the existence of this configure is provided in Ref.~\cite{Cho:1996qd}: $\pi_2(\mathbb{CP}^1) = \pi_2 (S^2) = \mathcal{Z}$. For the Higgs doublet $H=(H_1, H_2)^T$ with $H_1 H_1^* + H_2 H_2^* = v^2/2$, the vacuum manifold in the pure scalar sector is $S^3$ with $\pi_2(S^3)=0$. However, given the $U(1)_Y$ gauge freedom, one could make a gauge rotation to make one of the two complex fields $H_{1,2}$ real. As a result, the manifold has a lower dimension and is isomorphic to $S^2$. 
}
\beqa
H &=& \frac{v}{\sqrt{2}}\,\rho(r)\,\xi ~, \qquad \qquad  \qquad \qquad \xi = i 
\begin{pmatrix} 
\sin{(\frac{\theta}{2})} \,e^{-i\,\phi} \\ - \cos{(\frac{\theta}{2})} 
 \end{pmatrix} ~, \\
W^a_i &=& \epsilon^{aij} \,\frac{r^j}{r^2}\left( \frac{1- f(r)}{g}\right) ~, \qquad  \qquad 
W^a_0 = - \frac{v}{g}\, w(r)\, \frac{r^a}{r} ~,
\\
Y_i &=& - \frac{1}{g_{_Y}}( 1 - \cos{\theta})\, \partial_i \phi ~, \qquad \qquad  Y_0 = - \frac{v}{g_{_Y}}\,y(r) ~. 
\eeqa
Noting that $\xi^\dagger\, \vec{\sigma} \xi = - \vec{r}/r$, so $H^\dagger  \vec{\sigma} H$ has been treated as a triplet under $SU(2)_W$ as the simple $SU(2)$ monopole case~\cite{tHooft:1974kcl,Polyakov:1974ek}. Here, the index ``$i"$ for $W^a_i$ and $Y_i$ is the Cartesian coordinate index.  There are totally four dimensionless functions $\rho(r)$, $f(r)$, $w(r)$ and $y(r)$ to describe the Higgs and gauge field profiles. For the purely magnetic black hole case, one simply sets $w(r)=y(r)=0$. 
One can perform an $SU(2)_W$ gauge transformation to change from the hedgehog gauge to the unitary gauge
\beqa
\renewcommand{\arraystretch}{1.3}
\xi \longrightarrow U \xi = \begin{pmatrix} 
0 \\ 1
 \end{pmatrix} \qquad  \qquad
 \mbox{with} \qquad 
 U = - i\, \begin{pmatrix} 
\cos{(\frac{\theta}{2})} & \sin{(\frac{\theta}{2})}\,e^{-i \phi}  \\
\sin{(\frac{\theta}{2})}\,e^{i \phi} & -\cos{(\frac{\theta}{2})} 
 \end{pmatrix} ~. 
 \eeqa
In the unitary gauge and after rotating the neutral gauge fields from the basis $(Y_\mu, W^3_\mu)$ to the photon and $Z$ boson basis $(A_\mu, Z_\mu)$, one has
\beqa
\label{eq:photon}
A_\mu &=& - e\,v\, \left[ \frac{1}{g^2} \, w(r) + \frac{1}{g^2_{_Y}}\,y(r)\right]\, \partial_\mu t - \frac{1}{e} ( 1 - \cos{\theta}) \,\partial_\mu\, \phi ~, \\
\label{eq:zboson}
Z_\mu &=& \frac{e}{g\,g_{_Y}}\,v\, \left[ y(r) - w(r)\right] \, \partial_\mu t ~.
\eeqa
Note that $\partial_0 t = 1$ and $\partial_i t =0$. Again, for the purely magnetic black hole case with $w(r)=y(r)=0$, there is no $Z$ boson profile.

Substituting the ansatz profiles into the matter action, one has 
\beqa
S_{\rm matter} &\supset& \int d^4x \sqrt{- g} \,  \mathcal{L}_{\rm EW} \nonumber \\
&=& - 4\pi\, \int dt \, dr \,r^2\, 
\left[ P(r)\,N(r)\,\mathcal{K} + P(r)\,\mathcal{U} - P(r)^{-1}\,\mathcal{K}_0 - P(r)^{-1}N(r)^{-1}\,\mathcal{U}_0 \right] ~,
\eeqa
with 
\beqa
\mathcal{K} &=& \frac{v^2\,\rho'^2}{2} \,+\, \frac{f'^2}{g^2\,r^2} ~, \nonumber \\
\label{eq:Uformula}
\mathcal{U} &=& \frac{v^2\,f^2\,\rho^2}{4\,r^2} \,+\, \frac{(1-f^2)^2}{2\,g^2\,r^4}\, +\, \frac{\lambda}{8}\,v^4\,(\rho^2 - 1)^2\, + \, \frac{1}{2\,g_{_Y}^2\,r^4}\,\equiv \, \mathcal{U}_1 \,+\,\frac{1}{2\,g_{_Y}^2\,r^4} \, ~, \nonumber \\
\mathcal{K}_0 &=& \frac{v^2\,w'^2}{2\,g^2} \,+\, \frac{v^2\,y'^2}{2\,g_{_Y}^2} ~, \nonumber \\
\mathcal{U}_0 &=& \frac{v^2\,w^2\,f^2}{g^2\,r^2} \,+\, \frac{v^4\,(w-y)^2\,\rho^2}{8} ~.
\eeqa
The above formulas agree with Ref.~\cite{Cho:2016npz} for the magnetic case with $w=y=0$. Note that the term $1/(2\,g_{_Y}^2\,r^4)$ in $\mathcal{U}$ has infinite energy for the magnetic monopole without a black hole in the core~\cite{Cho:1996qd,Ellis:2016glu,Cho:2016npz}. This is another manifestation that the SM electroweak sector by itself does not admit a finite-energy magnetic monopole. Existence of a black hole event horizon can make the energy of the total system finite~\cite{Lee:1994sk,Weinberg:1994eb}. 

Variating the summed action $S_{\rm E} + S_{\rm matter}$ with respect to $P(r)$ and $N(r)$, the two Einstein equations are given by
\beqa
\label{eq:Nprime}
N' &=& \frac{1 - N}{r} - 8 \pi G\,r \left( \mathcal{U} + N\,\mathcal{K} + \frac{\mathcal{K}_0}{P^2}  + \frac{\mathcal{U}_0}{P^2\,N}   \right)  ~, \\
\label{eq:Pprime}
P' &=& 8 \pi G\,r\left(  P\, \mathcal{K} + \frac{\mathcal{U}_0}{P\,N^2} \right) ~.
\eeqa
Variating the action with respect to the matter fields $f(r)$, $\rho(r)$, $w(r)$ and $y(r)$, one has the following four matter equations of motion
\beqa
\label{eq:fprime}
\left(P\,N\,f' \right)' &=& P\left[ \frac{f(f^2 - 1)}{r^2} + \frac{g^2}{4}\,v^2\,f\,\rho^2 \right] - \frac{v^2\,f\,w^2}{P\,N}~, \\
\label{eq:rhoprime}
\left(r^2\,P\,N\,\rho' \right)' &=& \frac{1}{2}\,P\,\rho\,f^2 \,+\, \frac{\lambda\,v^2}{2}\,r^2\,P\,\rho(\rho^2 - 1) - \frac{v^2}{4\,P\,N} \,r^2\,\rho\,(w - y)^2 ~, \\
\label{eq:wprime}
\left(r^2\,P^{-1}\,w' \right)' &=& \frac{2}{P\,N}\,f^2\,w + \frac{g^2\,v^2}{4\,P\,N}\,r^2\,\rho^2\,(w - y) ~, \\
\label{eq:yprime}
\left(r^2\,P^{-1}\,y' \right)' &=& \frac{g_{_Y}^2\,v^2}{4\,P\,N}\,r^2\,\rho^2\,(y - w) ~.
\eeqa
%

\section{Hairy magnetic black holes}
\label{sec:magnetic}

\subsection{Masses and profiles}
\label{sec:magnetic-mass-profile}
For the magnetic black holes, existing papers have mainly studied  the $SU(2)$-gauge theory case~\cite{Lee:1991qs,Lee:1991vy,Maeda:1993ap,Lee:1994sk,Tachizawa:1994wn,Weinberg:1994eb}. Here, we focus on the SM electroweak $SU(2)_W \times U(1)_Y$ Lagrangian with our knowledge of the SM Higgs boson mass or the quartic coupling $\lambda$~\cite{Chatrchyan:2012ufa,Aad:2012tfa}. The equations of motion can be obtained from Eqs.~\eqref{eq:Nprime}\eqref{eq:Pprime}\eqref{eq:fprime}\eqref{eq:rhoprime} by setting $w(r)=y(r)=0$. We are looking for solutions with the existence of a horizon $r_H$. Defining 
\beqa
\label{eq:NrFr}
N(r) = 1 \, - \,\frac{2\,G\,F(r)}{r} + \frac{4\,\pi\,G}{g_{_Y}^2\,r^2} ~,
\eeqa
we have the asymptotic mass of the system to be $M=F(\infty)$. Substituting Eq.~\eqref{eq:Pprime} with $P'/P=8\pi G\,r\,\mathcal{K}$ into Eqs.~\eqref{eq:fprime}\eqref{eq:rhoprime}, we have three equations for three fields  $F(r)$ [or $N(r)$ via \eqref{eq:NrFr}], $f(r)$, $\rho(r)$
\beqa
\label{eq:Fprime} 
F' &=& 4\pi\,r^2 \left( \mathcal{U}_1 + N\,\mathcal{K} \right)  ~, \\
 \label{eq:fprime2}
\left( N \,f'\right)' + 8\pi\,G\,r\,N\,f'\,\mathcal{K} &=&  \frac{f(f^2 - 1)}{r^2} + \frac{g^2}{4}\,v^2\,f\,\rho^2~, \\
 \label{eq:rhoprime2}
  \left( r^2\,N \,\rho'\right)' +8\pi\,G\,r^3\,N\,\rho'\,\mathcal{K} &=&  \frac{1}{2}\,\rho\,f^2 \,+\, \frac{\lambda\,v^2}{2}\,r^2\,\rho(\rho^2 - 1) ~.
\eeqa
Noting that the $1/r^2$ term introduced in \eqref{eq:NrFr} is to have the equation of motion for $F(r)$ contain $\mathcal{U}_1$ without the last $1/r^4$ term in $\mathcal{U}$ [see \eqref{eq:Uformula}].

At the event horizon with $N(r_H)=0$, one has
\beqa
F(r_H) = \frac{r_H}{2\,G} + \frac{2\pi}{g_{_Y}^2\,r_H} \geq \frac{\sqrt{4\pi}}{\sqrt{G}\,g_{_Y}} = c_W\,\frac{\sqrt{4\pi}}{\sqrt{G}\,e} \equiv c_W\,M_{\rm eBH}^{\rm RN}~,
\eeqa
with $c_W \equiv \cos{\theta_W}$ and the extremal RN black hole mass $M_{\rm eBH}^{\rm RN}\equiv \sqrt{4\pi}\,M_{\rm pl}/e$.
Using \eqref{eq:NrFr}\eqref{eq:Fprime}, we obtain a boundary condition
\beqa
\label{eq:mag-BC1}
N' = \frac{1}{r} - 8 \pi\,G\,r\,\mathcal{U} \,, \qquad \mbox{at}~ r= r_H ~.
\eeqa
From \eqref{eq:fprime2}\eqref{eq:rhoprime2}, one has two more mixed BC's that are 
\beqa
\label{eq:mag-BC2}
&&N' \,f' =  \frac{f(f^2 - 1)}{r^2} + \frac{g^2}{4}\,v^2\,f\,\rho^2 \,, \qquad \mbox{at}~ r= r_H ~, \\
\label{eq:mag-BC3}
&&N' \,\rho' = \frac{1}{2}\,\frac{f^2\,\rho}{r^2} \,+\, \frac{\lambda\,v^2}{2}\,\rho(\rho^2 - 1)  \,, \qquad \mbox{at}~ r= r_H ~.
\eeqa
Together with the two BC's at infinity, $f(\infty) = 0$ and $\rho(\infty)=1$, there are totally five BC's that are required for the equations in \eqref{eq:Fprime}\eqref{eq:fprime2}\eqref{eq:rhoprime2}.

Before we solve these equations, we first discuss the ordinary RN black hole solution, for which $f(r)=0$ and $\rho(r) = 1$. Solving \eqref{eq:Fprime}, one has a simple solution for $F(r)$ as
\beqa
F(r) = M - \frac{2\pi}{g^2\,r} ~. 
\eeqa
With the solution of $P(r)=1$ to have an asymptotically flat metric, this matches the RN metric 
\beqa
P^2(r) N(r) = N(r) = 1- \, \frac{2 \,G\,M}{r} + \frac{4\pi G}{g^2 \,r^2} + \frac{4\pi G}{g_{_Y}^2 \,r^2} = 1- \, \frac{2 \,G\,M}{r} + \frac{4\pi\,G}{e^2 \,r^2} ~.
\eeqa
The outer horizon is at
\beqa
r_H \equiv r_+ = M\,G\,+\, \sqrt{M^2\,G^2 - 4\pi\,G/e^2}~,
\eeqa
provided that $M \geq M^{\rm RN}_{\rm eBH}=\sqrt{4\pi}M_{\rm pl}/e$. Inverting the above relation, one has 
\beqa
M^{\rm RN}_{\rm BH} = \frac{r_H}{2\,G} + \frac{2\pi}{e^2\,r_H} ~.
\eeqa
For the extremal case, one has $r^{\rm min}_H = r^{\rm min}_+ = \sqrt{4\pi}/(e\,M_{\rm pl})$.

Coming back to the hairy magnetic black holes, there exist constraints on the horizon $r_H$ if one makes a few plausible assumptions: $N'(r_H) \geq 0$, $f(r)$ is a monotonically decreasing function outside the horizon, while $\rho(r)$ is a monotonically increasing function. The requirement of $N'(r_H) \geq 0$ means $1/r_H > 8\pi\,G\,r_H\,\mathcal{U}(r_H)$ from \eqref{eq:mag-BC1}. For small $r_H$, one has $\mathcal{U}(r_H)\approx 1/(2g_{_Y}^2 r_H^4)$, so the lower bound on $r_H$ is 
\beqa
r_H \geq r_H^{\rm min} \equiv \frac{\sqrt{4\pi\,G}}{g_{_Y}} = c_W\,\frac{\sqrt{4\pi}}{e\,M_{\rm pl}}  ~,
\eeqa
which is smaller than the extremal RN black hole horizon radius by a factor of $c_W$. 
Applying those assumptions to the BC's in \eqref{eq:mag-BC2}\eqref{eq:mag-BC3}, one has 
\beqa
&&\lambda\,v^2\,r_H^2\,\left[1 - \rho^2(r_H)\right] < f^2(r_H) < 1 - \frac{g^2\,v^2\,r_H^2}{4}\,\rho^2(r_H) ~, \\
&&1 - \frac{f^2(r_H)}{\lambda\,v^2\,r_H^2} < \rho^2(r_H) < \frac{4}{g^2\,v^2\,r_H^2}\left[ 1- f^2(r_H)\right] ~.
\eeqa
For the SM with $g=e/s_W \approx 0.65$ and $\lambda=0.26$ such that $g^2 < 4 \lambda$ and using the fact that both $f(r_H)$ and $g(r_H)$ are within zero and one, the upper bound on $r_H$ is
\beqa
\label{eq:rH-upper}
r_H \leq \frac{2}{g\,v} = \frac{1}{m_W}\approx  2.5\times 10^{-3}\,\mbox{fm}~,
\eeqa
which is the characteristic radius of the monopole.

To calculate the mass of the hairy black holes, we integrate \eqref{eq:Fprime} to obtain
\beqa
F(r) = \int^r_{r_H} \, dr' \,e^{-\mathbb{K}(r', r)}\,4\pi\,r'^2 \left[ \mathcal{K}(r') + \mathcal{U}_1(r') + \frac{4\pi\,G}{g^2_{_Y}\,r'^2}\,\mathcal{K}(r')\, \right] + e^{-\mathbb{K}(r_H, r)}\,F(r_H) ~, 
\eeqa
where the new function is defined as
\beqa
\mathbb{K}(r', r) \equiv 8\pi\,G\,\int_{r'}^r\,dr''\,\mathcal{K}(r'') \,r'' ~.
\eeqa
Given that $G v^2 \ll 1$, the exponential, $e^{-\mathbb{K}(r', r)} = 1 + \mathcal{O}(Gv^2)$. Ignoring the terms equal to or higher than $\mathcal{O}(Gv^2)$, the mass of a hairy magnetic black hole has a simple formula
\beqa
\label{eq:MBH-mass}
M_{\rm hMBH}\,=\,F(\infty) &=& \int^\infty_{r_H} dr' 4\pi\,r'^2\,\left[ \mathcal{K}(r') + \mathcal{U}_1(r') \right] \,+ \, F(r_H) ~ \\
&=& \int^\infty_{r_H} dr' 4\pi\,r'^2\,\left[ \mathcal{K}(r') + \mathcal{U}_1(r') \right] \,+\,  \frac{r_H}{2\,G} + \frac{2\pi}{g_{_Y}^2\,r_H} ~.
\eeqa
The first integration term can be thought as the outside hair contribution to the total system mass. In the limit of $r_H \ll 1/m_W$, the integration is dominated by the region with $r' \sim 1/m_W$. One can then obtain the leading contribution by using a flat metric with $N(r)=1$. Ignoring terms proportional to $G$, the equations of motion for $f(r)$ and $\rho(r)$ are simply
\beqa
f'' &=& \frac{f(f^2 - 1)}{r^2} + \frac{g^2}{4}\,v^2\,f\,\rho^2 ~, \\
 \left( r^2\,\rho'\right)' &=& \frac{1}{2}\,\rho\,f^2 \,+\, \frac{\lambda\,v^2}{2}\,r^2\,\rho(\rho^2 - 1) ~,
\eeqa
which are similar to the t 'hooft-Polyakov $SU(2)/U(1)$ magnetic monopole case with a different $W$ mass in terms of $v$. Numerically solving the differential equations and then calculating the first integration term in \eqref{eq:MBH-mass}, the hairy magnetic black hole with $Q=2$ for $r_H \ll 1/m_W$ has a mass
\beqa
M_{\rm hMBH}&\approx& \frac{r_H}{2\,G} + \frac{2\pi}{g_{_Y}^2\,r_H}  \,+\,   0.75\times \frac{4\pi\,v}{g}=  \frac{r_H}{2\,G} + \frac{2\pi\,c_W^2}{e^2\,r_H} \,+\, 0.75\times \frac{2\pi\,v^2}{m_W}    \\
\label{eq:hMBH-hair-mass}
&\geq& c_W\,\frac{\sqrt{4\pi}\,M_{\rm pl}}{e} \,+\, 0.75\times \frac{2\pi\,v^2}{m_W}  = (1.2\times 10^{20} \,+\, 3.6\times 10^3) ~\mbox{GeV} ~.
\eeqa
Obviously, the hair part of the system contributes negligibly to the total mass. The upper mass of a $Q=2$ hairy magnetic black hole is 
\beqa
\label{eq:hMBH-mass-max}
M_{\rm hMBH} \leq M^{\rm max}_{\rm hMBH} = \frac{1}{2\,G\,m_W} + \mathcal{O}(m_W) \approx 9.3\times 10^{35}~\mbox{GeV}~. 
\eeqa

\begin{figure}[tb!]
	\label{fig:mag-extremal}
	\begin{center}
		\includegraphics[width=0.48\textwidth]{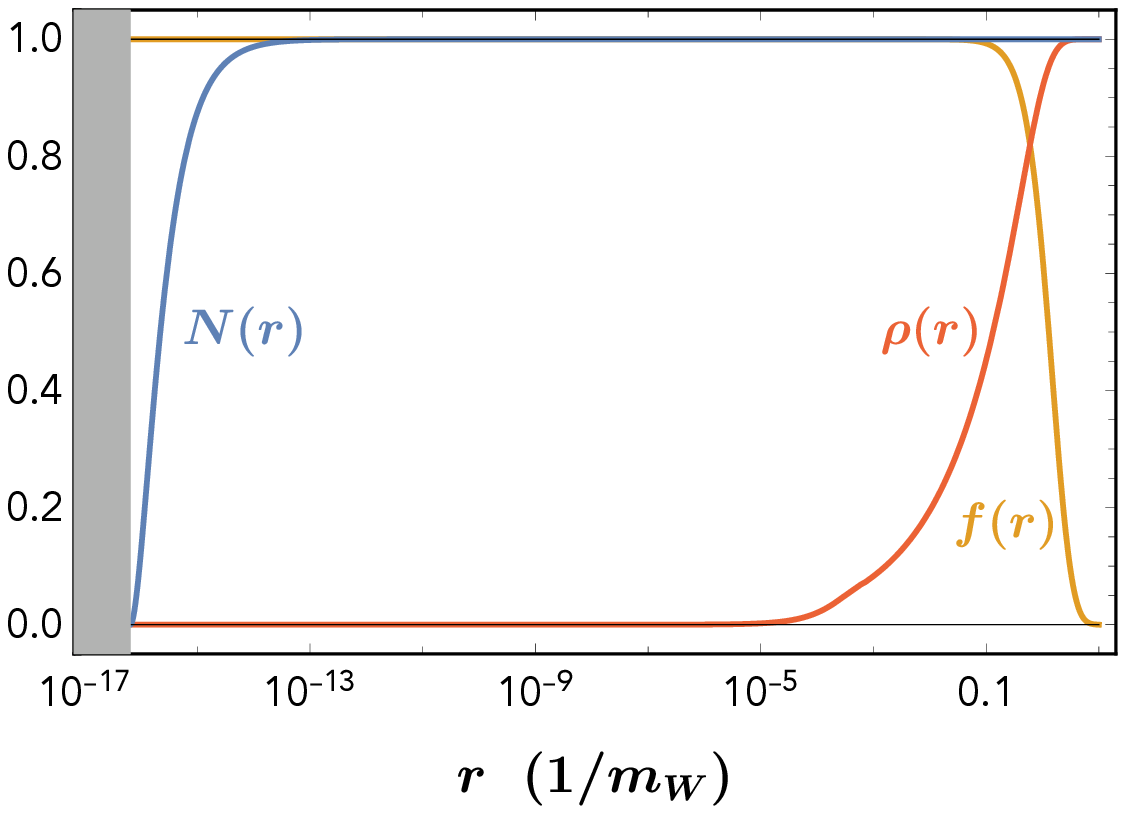} \hspace{3mm} 
		\includegraphics[width=0.48\textwidth]{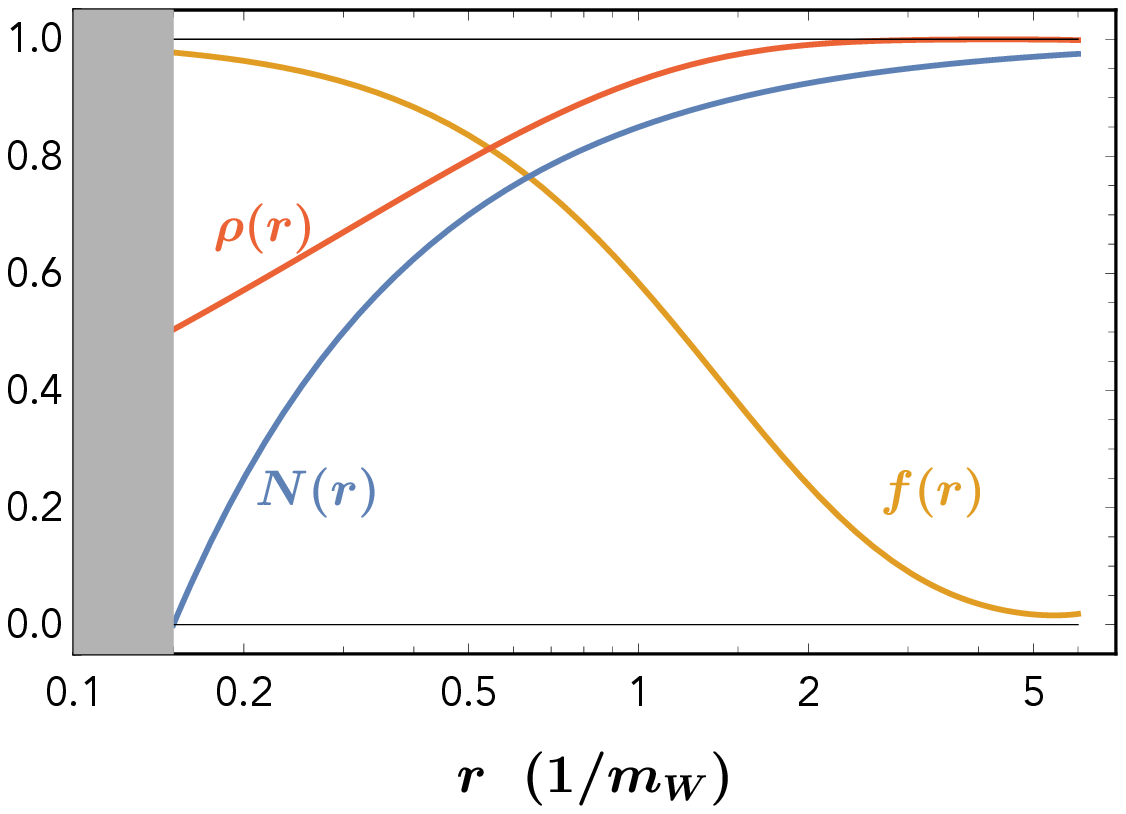}
		\caption{Profiles as a function of $r$ for the SM hairy magnetic black holes. {\it Left:} the horizon $r_H =  c_W\sqrt{4\pi}/(e\,M_{\rm pl})$ for an extremal black hole. {\it Right:} a larger horizon with $r_H = 0.15/m_W$ with the Higgs vacuum expectation value (VEV) profile $\rho(r)$ to be half-restored at the horizon. Not shown here is the profile for $P(r)$, which is approximately one because of $P'/P = \mathcal{O}(G v^2)$ and $P(\infty) =1$.
		 }
	\end{center}
\end{figure}

\begin{figure}[tb!]
	\label{fig:mass-radius}
	\begin{center}
		\includegraphics[width=0.48\textwidth]{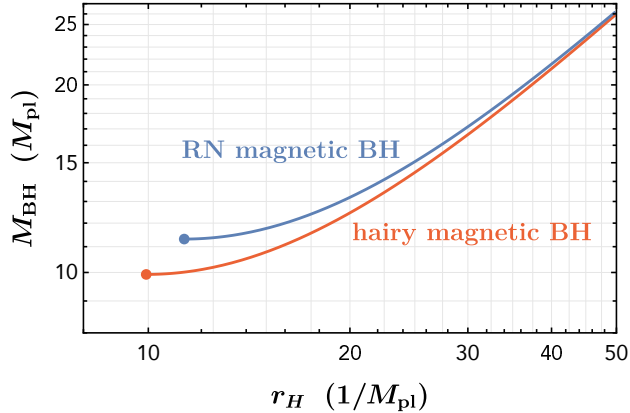} 
		\caption{The black hole masses as functions of horizon radii for both the hairy and RN magnetic black holes. The left endpoints of the two curves correspond to the extremal black hole cases. 
		 }
	\end{center}
\end{figure}

Numerically solving the equations of motion with the BC's, we show two representative profiles in Fig.~\ref{fig:mag-extremal}. In the left panel with $r_H = r_H^{\rm min}$ corresponding to the extremal case, the black hole sits well inside the hairy cloud. Around the event horizon, the Higgs VEV is very close to zero and the electroweak symmetry is almost completely restored.~\footnote{This is subject to corrections from the QCD condensation induced electroweak symmetry breaking.} For potentially phenomenological applications, we also provide numerically fitted functions for both $f(r)$ and $\rho(r)$, that are good approximation for $r > 0.1\,m_W^{-1}$,
\beqa
f(r) &\approx& \frac{0.495\,m_W\,r}{\sinh{(1.1\,m_W\,r})} + \frac{1.265\,m_W\,r}{\sinh{(2.3\,m_W\,r})} ~, \\
\rho(r) &\approx& \left(\coth{\left[86\,(m_W\,r)^{1.2}\right]} - \frac{1}{86\,(m_W\,r)^{1.2}} \right) \left( 1 - 0.51\,e^{-1.82\,m_W\,r} \right) ~.
\eeqa
In the right panel of Fig.~\ref{fig:mag-extremal}, we show an example of the non-extremal case with $r_H = 0.15\,m_W^{-1}$ and the Higgs VEV half-restored at the event horizon. The total mass of system is well approximated by the term $r_H/(2 \, G)$ in Eq.~\eqref{eq:MBH-mass}. The hair mass is 3.4~TeV and smaller than the extremal case in \eqref{eq:hMBH-hair-mass}. In Fig.~\ref{fig:mass-radius}, we show black hole masses as a function of horizon radii. For the same mass, the hairy magnetic black hole has a larger horizon radius, hence more entropy, than the RN magnetic black hole.

\subsection{Hawking temperature and black hole evaporation}
\label{sec:temperature-evolution}
A non-extremal black hole evaporates via the Hawking radiation process.~\footnote{If there exist monopoles in grand unified theories, the Schwinger discharge effect will evaporate away the magnetic charge of an hMBH. Based on SM interactions alone, there are no finite-energy monopoles, so the possibility of the Schwinger discharge effect is ruled out.} Based on the black hole thermodynamics, the black hole temperature is proportional to the surface gravity $\kappa$ at the horizon. For a Killing vector $\xi^a$, the surface gravity is $\kappa^2 = -(\nabla^a \xi^b )(\nabla_a \xi_b)/2$ evaluated at the horizon~\cite{Wald:106274}. Using the metric in \eqref{eq:metric}, we have
\beqa
T = \frac{\kappa}{2\pi}  = \frac{P(r_H) N'(r_H)}{4\pi} ~.
\eeqa
The factor $P(r_H) = \mbox{exp}[- \int^\infty_{r_H} dr\,8\pi\,G\,r\,\mathcal{K}(r) ] = 1 \,+\,\mathcal{O}(Gv^2)$. The other factor $N'(r_H)$ can be obtained using \eqref{eq:mag-BC1}: $N'(r_H) = 1/r_H - 8\pi\,G\,r_H\,\mathcal{U}(r_H) =1/r_H - 4\pi\,G/(g_{_Y}^2\,r_H^3) + \mathcal{O}(G v^2/r_H)$. So, we have
\beqa
T = \frac{1}{4\pi\,r_H} \left( 1- \frac{4\pi\,G}{g_Y^2\,r_H^2}\right) = \frac{M_{\rm pl}^2}{2\pi}\,\frac{\sqrt{M^2 - M^2_{\rm eBH}} }{\left( M + \sqrt{M^2 - M^2_{\rm eBH}}\right)^2} ~,
\eeqa
where the extremal hairy black hole mass is $M_{\rm eBH} \equiv c_W\sqrt{4\pi}M_{\rm pl}/e$. The above temperature formula is similar to the RN black hole case, except that the extremal mass is smaller by a factor of $c_W$. This factor can be thought as the electroweak symmetry restoration near the event horizon, so the black hole carries magnetic hypercharge $2\pi Q/g_Y = c_W 2\pi Q/e$ with $Q=2$. Note that $T$ is not a monotonic function of $M$. For a very heavy mass $M \gg M_{\rm eBH}$, it is suppressed by $1/M$ like the Schwarzschild case. For the opposite limit with $M \rightarrow M_{\rm eBH}$, $T \propto \sqrt{M - M_{\rm eBH}}$ and also suppressed. The maximal Hawking temperature has 
\beqa
T^{\rm max} = \frac{e\,M_{\rm pl}}{12\sqrt{3}\,\pi^{3/2}\,c_W} \approx 3.8\times 10^{16}\,\mbox{GeV}~,
\eeqa
when $M= 2 M_{\rm eBH}/\sqrt{3}$. 

Depending on the early universe production mechanism, the initial state could be either a RN magnetic black hole or a hairy one. For the former case, the Hawking radiation leads to shrinking of $r_H$. When $r_H < 1/m_W$ or $M<M^{\rm max}_{\rm hMBH}$ in \eqref{eq:hMBH-mass-max}, the RN black hole is not classically stable and quickly converts to the hairy magnetic black hole~\cite{Lee:1991qs}. At this moment, the temperature is around 6.4~GeV. The black hole horizon continues shrinking and $T$ keeps increasing. During this period, the Hawking radiation is dominated by the $2d$ one (the ground state of charged fermions in the monopole background behaves as $2d$ modes) with the radiated power as~\cite{Maldacena:2020skw,Bai:2020spd} 
\beqa
P_2 = -\frac{dM}{dt} = \frac{\pi\,g_*}{24}\,T^2(M, M_{\rm eBH}) ~,
\eeqa
with $g_*$ counts the number of left- and right-handed $2d$ modes using the hypercharges of chiral fermions: $g_*=|Q|$ for $q_L$, $\ell_L$, $d_R$, $e_R$ and $g_*=2|Q|$ for $u_R$. Before the black hole mass becomes close to the extremal mass or in the limit of $M \gg  M_{\rm eBH}$, it evaporates with a time scale of 
\beqa
\label{eq:tau-I}
\tau_{\rm I} \approx \frac{512\pi M^3}{g_* \, M_{\rm pl}^4} = (1.0\times 10^9\,\mbox{s})\times (M/M^{\rm max}_{\rm hMBH})^3 ~,
\eeqa
for $g_*=36$. So, for $M  \gtrsim 10^{-3} M^{\rm max}_{\rm hMBH} \approx 10^{33}$~GeV, this period of evaporation is longer than the Big Bang nucleosynthesis (BBN) time $\mathcal{O}(1\,\mbox{s})$. The black hole evaporation could leave observational imprints on the light-element abundance~\cite{Bai:2019zcd}. 

As the black hole approaches the extremal state, the Hawking temperature is approximately $T \simeq M_{\rm pl}^2 \sqrt{M - M_{\rm eBH}}/(\sqrt{2}\pi\,M_{\rm eBH}^{3/2})$. For $T \gtrsim m_e$ [$M-M_{\rm eBH} \gtrsim 16\pi^{7/2} c_W^3\,m_e^2/(e^3\,M_{\rm pl})\approx 4.1\times 10^{-22}\,\mbox{GeV}$] or the lightest electrically-charged particle mass, $2d$ radiation is still the dominant one. The evaporation time scale from some initial mass $M\simeq M_{\rm eBH}$ to the mass with $T \simeq m_e$ is estimated to be
\beqa
\tau_{\rm II} \,=\, \frac{384\,\pi^{5/2}\,c_W^3}{e^3\,g_*\,M_{\rm pl}} \ln{\left[ \frac{M_{\rm pl}^4\,(M-M_{\rm eBH})}{2\pi^2\,m_e^2\,M_{\rm eBH}^3} \right]} \approx 3.7 \times 10^{-37}~\mbox{s}~,
\eeqa
for $M = 1.1 \,M_{\rm eBH}$ and $g_*=2$. This period of $2d$ evaporation is very short. 

As the black hole temperature continues to decrease till $T \lesssim m_e$, the $2d$ evaporation is not active any more. The ordinary $4d$ blackbody radiation has
\beqa
P_4 = -\frac{dM}{dt} \approx \frac{\pi^2\,g_*}{120}\,4\pi R_{\rm mono}^2\, T^4(M, M_{\rm eBH}) ~.
\eeqa
Here, $g_* = 2$ for photon and $g_* = 21/4$ for three chiral neutrinos; $R_{\rm mono} \approx 1/m_W$ is the monopole hair size, within which both photons and neutrinos are stored assuming that the hairy part is heated up to the Hawking temperature of the black hole. Solving the above equation, we have the black hole mass approaching the extremal state as
\beqa
M \,\approx \, M_{\rm eBH} \,+\,  \frac{120\pi\,M_{\rm eBH}^6\,m_W^2}{g_*\,M_{\rm pl}^8}\, \frac{1}{t} ~. 
\eeqa
During this long process, the Hawking temperature is
\beqa
T \approx \sqrt{\frac{60}{\pi\,g_*}}\, \frac{m_W\,M_{\rm eBH}^{3/2}}{M_{\rm pl}^2}\, t^{-1/2}  \,\approx \, 2.7\times 10^{-27}\,\mbox{GeV} \times \left( \frac{t_{\rm univ}}{t} \right)^{1/2} ~, 
\eeqa
with $g_* = 2$ and the age of the Universe $t_{\rm univ} \approx 4.35\times 10^{17}$~s. After the initial evaporation time $t_{\rm I}$ in \eqref{eq:tau-I}, the Hawking temperature is already suppressed to be smaller than $\sim 10^{-23}$~GeV, which is tiny. 

\section{Hairy dyonic black holes}
\label{sec:dyonic}

\subsection{Masses and profiles}
\label{sec:dyonic-mass-profile}
For the dyonic black hole case, we have six equations of motion in 
Eqs.~\eqref{eq:Nprime}-\eqref{eq:yprime} with four second-order-differential and two first-order-differential equations. We need $2\times 4 + 2 = 10$ BC's to solve the system. Four BC's are the mixed boundary conditions and can be obtained by simply evaluating \eqref{eq:fprime}-\eqref{eq:yprime} at $r=r_H$. Noting that $N(r_H) = 0$ and to have a non-divergent BC's, two of these four BC's become simple conditions for $w(r)$ and $y(r)$ as~\cite{Brihaye:1998cm,Brihaye:1999nn}
\beqa
w(r_H) = 0 \,, \qquad y(r_H) = 0 ~. 
\eeqa
Two more BC's are the existence of the event horizon and the asymptotically flat spacetime at infinity and four more BC's are the profiles at infinity
\beqa
N(r_H) = 0 \,,\quad P(\infty) = 1 \,,\quad  f(\infty) =0 \,, \quad \rho(\infty) = 1 \,, \quad w(\infty) = y(\infty) = w_\infty ~.
\eeqa
Noting from \eqref{eq:photon}\eqref{eq:zboson},  $w(\infty) = y(\infty)$ can assure vanishing $Z$ profile at infinity, but not the photon profile.

Similar to the magnetic case, we define the function $F(r)$ via \eqref{eq:NrFr}, and \eqref{eq:Nprime} can be replaced by 
\beqa
\label{eq:Fprime-dyon} 
&&F' = 4\pi\,r^2 \left(\mathcal{U}_1 +  N\,\mathcal{K} + \frac{\mathcal{K}_0}{P^2} + \frac{\mathcal{U}_0}{P^2\,N}\right) ~.
\eeqa
The boundary condition $N(r_H) = 0$ can be converted to 
\beqa
F(r_H) \,=\, \frac{r_H}{2\,G} \,+\, \frac{2\pi}{g_{_Y}^2\, r_H} ~,
\eeqa
which is similar to the magnetic case. At $r=r_H$, the derivative $N'(r_H)$ is given by 
\beqa
\label{eq:Np-rH-dyon}
N'(r_H) = \frac{1}{r_H} - 8 \pi\, G\, r_H \left. \left( \mathcal{U} + \frac{\mathcal{K}_0}{P^2}  + \frac{\mathcal{U}_0}{P^2\,N} \right)\right|_{r_H} ~.
\eeqa

The EOMs admit the ordinary RN dyonic black hole solutions with $f(r)=0$, $\rho(r)=1$, $P(r)=1$ and $\omega(r)=y(r)$.  For this solution, one has $(r^2\,y')' = 0$.
Subject to the two BC's, $y(r_H)=0$ and $y(\infty)=y_\infty$, the solution is 
\beqa
\label{eq:y-solution}
w(r) =y(r) = y_\infty \left( 1 - \frac{r_H}{r} \right) ~. 
\eeqa
From \eqref{eq:photon}, the zeroth component of the photon field in the unitary gauge is 
\beqa
A_0 = -e\,v \,\left( \frac{1}{g^2} + \frac{1}{g_{_Y}^2} \right)\,y(r) = - \frac{v}{e}\, y(r) ~.
\eeqa
The asymptotic radial electric field is 
\beqa
E_r  = F_{0r} = - \partial_r A_0 =  \frac{v}{e}\,y'(r)  = \frac{v\,r_H\,y_\infty}{e\,r^2} \equiv \frac{q\,e}{4\pi\,r^2}~,
\eeqa
from which, one can identify $y_\infty = q\,e^2/(4\pi\,v\,r_H)$. The equation of motion for $F(r)$ becomes
\beqa
F' = \frac{2\pi}{g^2\,r^2}\,\left( 1 + \frac{g^2}{e^2}\,y_\infty^2\,v^2\,r_H^2\right) =  \frac{2\pi}{g^2\,r^2}\, \left[ 1 + \frac{q^2\,g^2\,e^2}{(4\pi)^2} \right] ~.
\eeqa
The solution to the above equation is
\beqa
F(r) = M - \frac{2\pi}{g^2\,r} \left[ 1 + \frac{q^2\,g^2\,e^2}{(4\pi)^2} \right]~. 
\eeqa
Substituting this into \eqref{eq:NrFr}, we have the RN metric
\beqa
P^2(r) N(r) = N(r) =  1- \, \frac{2 \,G\,M}{r} + \frac{4\pi\,G}{e^2 \,r^2} + \frac{G\,q^2\,e^2}{4\pi\,r^2} 
= 1- \, \frac{2 \,G\,M}{r} + G\, \frac{Q^2\,e_{\rm M}^2 \,+ \,q^2\,e^2}{4\pi\,r^2} 
 ~,
\eeqa
which matches the RN black hole with the magnetic and electric charge of $(Q=2, q)$ and the magnetic coupling of $e_{\rm M} = 2\pi/e$. The extremal dyonic RN black hole mass and horizon radius are 
\beqa
\label{eq:RN-dyon-extremal}
M^{\rm RN}_{\rm eBH}=\frac{\sqrt{4\pi}M_{\rm pl}}{e}\, \left[ 1 + \frac{q^2\,e^4}{(4\pi)^2}  \right]^{1/2} ~, \qquad 
r^{\rm min}_H = r^{\rm min}_+ = \frac{\sqrt{4\pi}}{e\,M_{\rm pl}} \, \left[ 1 + \frac{q^2\,e^4}{(4\pi)^2}  \right]^{1/2} ~.
\eeqa

Coming back to the hairy dyonic black holes, we first work out the constraints on $r_H$. The lower limit can be obtained by requiring $N'(r_H) \geq 0$ from \eqref{eq:Nprime}. In the limit of $Gv^2 \ll 1$, $P(r) \approx 1$ and $N(r) \approx ( 1- r_H/r)^2$ for the extremal case. In the limit of $r\rightarrow r_H$, one has $y(r) \propto r_H^{-1}(r - r_H)$, $w(r) \propto m_W\,r_H^{-1}( r - r_H)^2$, $f(r) \approx 1$ and $\rho(r) \propto ( 1- r_H/r)^{(\sqrt{3}-1)/2}$. The leading contribution to $\mathcal{K}_0\approx v^2\,[y'(r_H)]^2/(2 g_{_Y}^2)$ is $\mathcal{O}(1/r^2_H)$ instead of $\mathcal{O}(1/r^4_H)$ for the RN black hole case. Therefore, keeping the dominant $1/(r^4_H)$ term in $\mathcal{U}$ for \eqref{eq:Np-rH-dyon} and in the limit of $Gv^2 \ll 1$, we have 
\beqa
N'(r_H) \approx \frac{1}{r_H} - 8\pi G\, r_H \left( \frac{1}{2\,g_{_Y}^2\,r_H^4} \right) \geq 0~.
\eeqa
From the above condition, one has the hairy black hole horizon above the extremal one
\beqa
\label{eq:dyon-rH-min-1}
r_H \geq r_H^{\rm min} = \frac{\sqrt{4\pi\,G}}{g_{_Y}} = c_W\,\frac{\sqrt{4\pi}}{e\,M_{\rm pl}}  ~,
\eeqa
which is the same as the hairy magnetic black hole case.  One may wonder why there is not any additional $q$-dependent correction, which is present for the extremal RN dyonic black hole case in \eqref{eq:RN-dyon-extremal}. This can be understood by looking at the behaviors of \eqref{eq:yprime} in $r$, since $r^2\, y'$ is effectively the amount of hypercharge as a function of $r$. At large $r$, one has $y(r) \approx w(r)$,  $r^2\, y'$ stays constant and the effective charge does not change much. For a smaller $r$ close to the scale of $r_H$ and because of the double zero in $N(r)$, the positive slope of the effective charge in $r$ becomes very large. As a result, the effective charge decreases dramatically as $r$ decreases. As $r$ is very close to $r_H$, the effective charge becomes approximately zero. So, the extremal hairy dyonic black hole has a magnetic black hole in the core and dyonic hair outside. 

For the upper bound on $r_H$, noting that both $w(r) \propto (r - r_H)$ and $y(r) \propto (r - r_H)$ and $N \propto (r - r_H)$ when $r \rightarrow r_H$ for a non-extremal black hole, so the last terms in \eqref{eq:fprime} and \eqref{eq:rhoprime} can be ignored. Requiring $f'(r_H) \leq 0$ and $\rho'(r_H) \geq 0$, one can derive the same upper limit $r_H \leq 1/m_W$ as in \eqref{eq:rH-upper} for the hairy magnetic black hole case. Hence, one has a similar upper mass limit for hairy dyonic black holes $M_{\rm hDBH}^{\rm max}\approx 9.3\times 10^{35}$~GeV as in \eqref{eq:hMBH-mass-max}. 

Based on \eqref{eq:fprime} and requiring an exponentially decreasing behavior for $f(r)$ at large $r$, the electric charge is bounded from above~\cite{Julia:1975ff}. The constraint on the large-$r$ value $y_\infty$ of $y(r)$ to be $y_\infty \leq g/2$, which could be converted to an upper bound on the electric charge $q$. Although an analytic result can not be obtained, one could make some approximation to estimate the magnitude of the upper bound on $q$. To proceed, we make a step-function assumption for $f(r)$ such that $f(r) = 1 (0)$ for $r \leq(>)\,m_W^{-1}$. In the large $r$ limit and from \eqref{eq:yprime} and \eqref{eq:wprime}, one has $y(r)= y_\infty - b_1/r$ and $w(r) = y_\infty - b_2/r$.  Matching the electric charge, one has $b_1 \sim b_2 \sim b = q\,e^2/(4\pi\,v)$. In the small $r$ limit and noting that $N(r)\approx (1 - r_H/r)$ for non-extremal cases, one has  $y(r) \propto r_H^{-1}(r - r_H)$ and  $w(r) = c_2 (r - r_H)$. Requiring both $w(r)$ and $w'(r)$ to be continuous at $r=m_W^{-1}$, one has $b\approx m_W^{-1}\,y_\infty/2$. The condition of $y_\infty \leq g/2$ becomes $q \lesssim 2\pi/e^2$, which says that the electric charge times the electric coupling can not be larger than the magnetic charge times the magnetic coupling to have a hairy dyonic black hole. There is no such requirement for the ordinary RN dyonic black holes. 

\begin{figure}[th!]
	\label{fig:dyon-mass-charge}
	\begin{center}
		\includegraphics[width=0.5\textwidth]{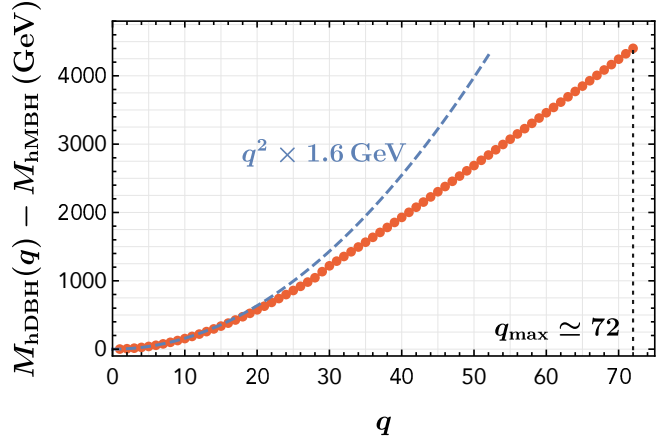} 
		\caption{The mass difference between extremal hairy dyonic and magnetic black holes for different electric charge $q$. The upper charge limit $q_{\rm max}\simeq72$ or $\mathcal{O}(2\pi/e^2)$ is obtained by requiring an exponentially decreasing behavior for $f(r)$ at large $r$. 
		 }
	\end{center}
\end{figure}

In the limit of $Gv^2 \ll1$ and integrating \eqref{eq:Fprime-dyon}, we have the mass of a hairy dyonic black hole as
\beqa
M_{\rm hDBH} &=& F(\infty) = \int^\infty_{r_H} dr' 4\pi\,r'^2\,\left[  \mathcal{U}_1(r') + N\,\mathcal{K}(r') + \frac{\mathcal{K}_0(r')}{P^2} + \frac{\mathcal{U}_0(r')}{P^2\,N} \right] \,+\,F(r_H) \nonumber \\
&\approx& \int^\infty_{\sim 10\,r_H} dr' 4\pi\,r'^2\,\left[  \mathcal{U}_1(r') + \mathcal{K}(r') + \mathcal{K}_0(r')  + \mathcal{U}_0(r') \right] \,+\,\frac{r_H}{2\,G} \,+\, \frac{2\pi}{g_{_Y}^2\, r_H}  ~.
\eeqa
where in the second line the hair mass can be calculated by treating the spacetime as a flat one. In the limit of $r_H \ll m_W^{-1}$, the equations of motion for $f(r)$, $\rho(r)$, $w(r)$ and $y(r)$ can be solved with $N(r) \approx 1$ and $P(r)\approx 1$. For a small electric charge with $q \ll 2\pi/e^2$, one can first ignore the last terms in \eqref{eq:fprime} and \eqref{eq:rhoprime} and obtain the solution for $f(r)$ and $\rho(r)$. Substituting them into \eqref{eq:wprime} and \eqref{eq:yprime}, one can solve just two coupled differential equations to obtain $w(r)$ and $y(r)$. We have also solved the full-coupled equations to check that this approximation is good. For $q \ll 2\pi/e^2$, a simple formula for the mass in terms of $q$ is given by
\beqa
M_{\rm hDBH}(q) &\approx&  \frac{r_H}{2\,G} + \frac{2\pi\,c_W^2}{e^2\,r_H}\, + \,  (0.75 + 0.00034\,q^2)\times \frac{2\pi\,v^2}{m_W} \nonumber \\
&\approx& M_{\rm hMBH} + q^2\,\times\,1.6~\mbox{GeV} ~. 
\eeqa
At the order-of-magnitude level, the mass difference between the dyonic and magnetic hairy parts is $M_{\rm hDBH} - M_{\rm hMBH}= \mathcal{O}(q^2\times \alpha\,m_W)$ with $\alpha$ as the fine-structure constant. Solving the fully coupled equations, in Fig.~\ref{fig:dyon-mass-charge} we show the mass difference as a function of charges for the extremal black hole case. As discussed before, there exists an upper limit on the charge $q_{\rm max} \simeq 72$ (up to some numeric uncertainties) for the extremal case to have $w_\infty=y_\infty < g/2$ or to have an exponentially decreasing behavior for $f(r)$ at large $r$. As one can see from this plot, the $q^2$ behavior matches very well the numerically calculated results for small $q$.

\begin{figure}[t]
	\label{fig:dyon-extremal-profile}
	\begin{center}
		\includegraphics[width=0.48\textwidth]{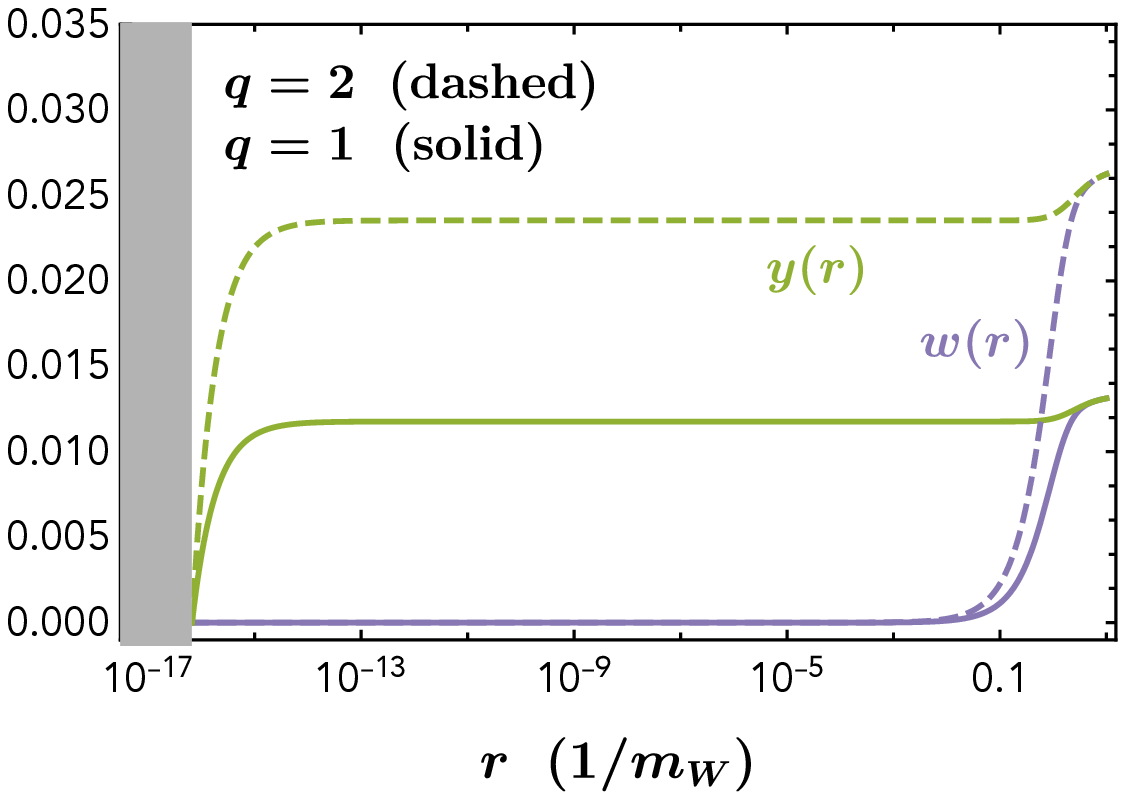} \hspace{3mm} 
		\includegraphics[width=0.455\textwidth]{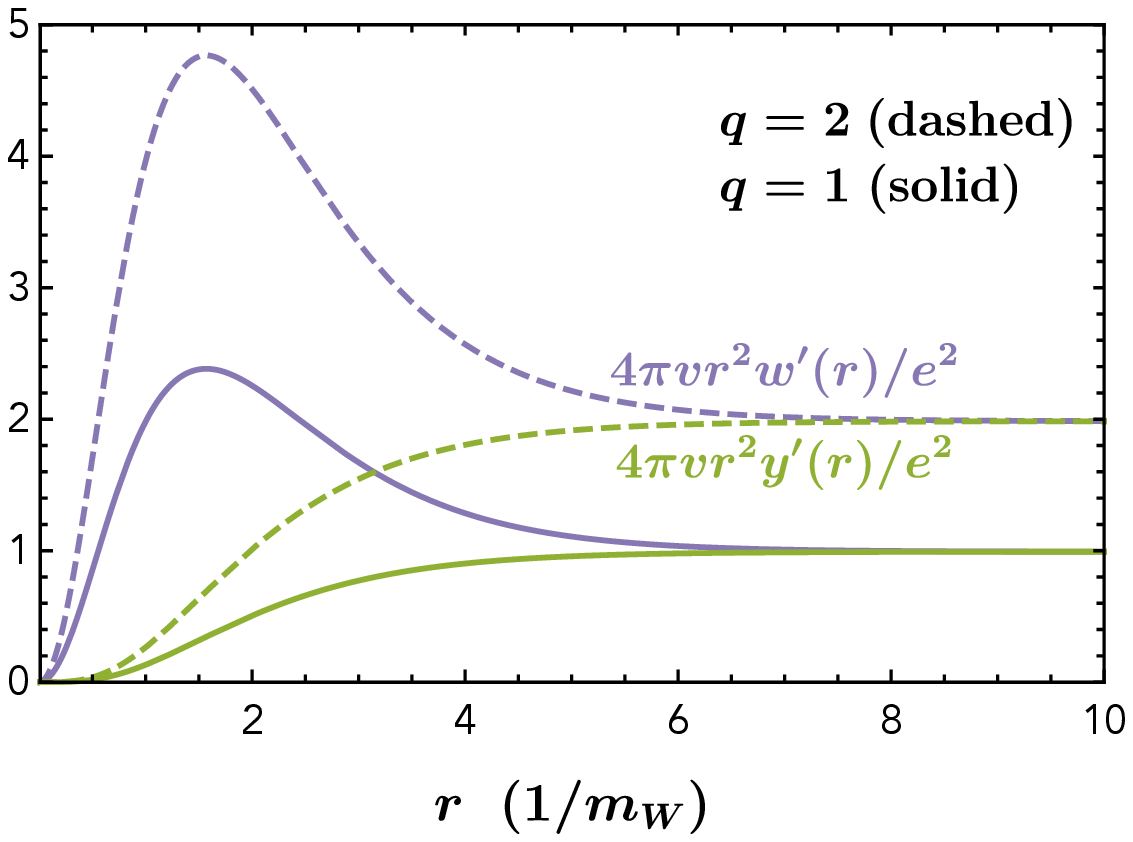}
		\caption{ {\it Left:} $y(r)$ and $w(r)$ as a function of $r$ for the SM hairy dyonic black holes with an extremal black hole horizon $r_H =  c_W\sqrt{4\pi}/(e\,M_{\rm pl})$. The profiles for other functions are approximately the same as in the left panel of Fig.~\ref{fig:mag-extremal}.  {\it Right:} The effective charges related to the two neutral gauge fields as a function of $r$. 
		 }
	\end{center}
\end{figure}

In the left panel of Fig.~\ref{fig:dyon-extremal-profile}, we show the profiles $y(r)$ and $w(r)$ for the extremal hairy dyonic black hole with $q=1$ and $q=2$. For $y(r)$, it increases very fast from zero at $r = r_H$ to a plateau at $r \sim 10\,r_H$. At $r \sim m_W^{-1}$, it continuously increases by a small amount into an asymptotic value $y_\infty$ at $r \gg m_W^{-1}$. For $w(r)$, it stays almost zero until $r \sim m_W^{-1}$, where it starts to increase to its asymptotic value $\omega_\infty = y_\infty$. For $r \gg m_W^{-1}$ and for $q \ll 2\pi/e^2$, one has the following asymptotic behaviors
\beqa
w(r) &=& q\,\frac{e^2\,m_W}{4\pi\,v}\,\left( 5.2 \,-\, \frac{1}{m_W\,r} \,-\,6.7\,c_W^2\,\times \frac{e^{-m_Z\,r}}{m_W\,r}  \right)~, \\
y(r) &=& q\,\frac{e^2\,m_W}{4\pi\,v}\,\left( 5.2 \,-\, \frac{1}{m_W\,r} \,+\,6.7\,s_W^2\,\times \frac{e^{-m_Z\,r}}{m_W\,r} 
\right) ~.
\eeqa
Using \eqref{eq:photon} and \eqref{eq:zboson}, one can also obtain the asymptotic behaviors of the time components of photon and $Z$-boson fields and again in the small $q$ limit
\beqa
\label{eq:A0r}
A_0(r) &\xrightarrow{r\rightarrow \infty}& - q\,\frac{e \,m_W}{4\pi}\,\left( 5.2 \,-\, \frac{1}{m_W\,r}  \right) ~, \\
Z_0(r) &\xrightarrow{r\rightarrow \infty}& 0.07\times q\,\frac{e^{-m_Z\,r}}{r}  ~.
\eeqa
Because of the exponentially decreasing dependence for the time component of the $Z$-boson field, the hairy dyonic black hole has only a long-range electric field outside the hair radius of $\sim m_Z^{-1} \sim m_W^{-1}$.

\subsection{Some properties of hairy dyonic black holes}
\label{sec:dyon-properties}
Since the core region of a hairy dyonic black hole is only a magnetic black hole with electroweak symmetry restoration just outside the horizon. The black hole mass evolution as a function of time is the same as discussed in Section~\ref{sec:temperature-evolution}. On the other hand, the existence of an electric charge for a dyonic black hole makes the hairy dyonic black holes different from the magnetic one. Here, we will focus on the extremal case with $r_H = r_H^{\rm min}$ in \eqref{eq:dyon-rH-min-1}. 

From Fig.~\ref{fig:dyon-mass-charge}, the mass difference between two hairy dyonic black holes with $\Delta q =1$ grows as $q$ increases. The largest difference is $M_{\rm hDBH}(72) - M_{\rm hDBH}(71)\approx 79.5$~GeV, which is close to but below the free $W$-gauge boson mass. So, a charge-$q$ hDBH can not decay into a charge-$q$-1 hDBH plus an on-shell $W$-gauge boson. On the other hand, it can be discharged by generating light charged fermions based on the charged current in the SM. Such processes have been studied and calculated based on the Julia-Zee dyons in the large charge limit~\cite{Blaer:1981ps,Blaer:1981ui,Callan:1982ah} . For the $q=1$ case, it will decay into a hairy magnetic black hole plus kinematically-accessible charged particles in the SM
\beqa
\mbox{hDBH}^- \rightarrow \mbox{hMBH} + \{ \pi^- \,, \pi^- \, \pi^0 \,, \pi^- \pi^- \pi^+ \,,  \mu^- \, \bar{\nu}_\mu \,, e^-\, \bar{\nu}_e \,, \cdots \}\,, 
\eeqa
which is similar to the $\tau$-lepton decays since the mass difference is around 1.6~GeV and close to the  $\tau$-lepton mass. Different from the $\tau$ lepton, the dyon decay width is not suppressed by two powers of the Fermi constant. Based on a classical field theory calculation, its decay width is estimated to be~\cite{Sonoda:1983qh}
\beqa
\Gamma(\mbox{hDBH}^- ) \sim \frac{\alpha}{4\pi}\,m_W ~. 
\eeqa
So, an hDBH decays very fast into an hMBH state on the time scale of $10^{-23}$~s. 

For a non-integer-charged hDBH, the discharge processes can reduce its electric charge to be below one. After that, the hDBH with $q < 1$ is also a cosmologically stable state. However, just based on the SM with all color-singlet states having integer charges, it is unlikely to form a non-integer-charged hDBH. One exception is from a non-zero CP-violating $\theta$-term, $\theta\,\alpha/(8\pi)\,F_{\mu\nu}\widetilde{F}^{\mu\nu}$, in the SM Lagrangian. There is no zero-charged hMBH because of the $\theta$-dependent quantization condition~\cite{Witten:1979ey}. The electric charge of hDBH is $n - \theta/(2\pi)$ with $n$ as an integer. For $n=0$, the irrational-charged hDBH could be a cosmologically stable object. For $|\theta| > 8\pi^2 m_e^2/(e^2 \, m_W^2) \approx 3.2\times 10^{-8}$, the electric field at the monopole radius $\sim m_W^{-1}$ is above $m_e^2/e$, so additional electron and positron clouds should also exist as additional hair for the black hole~\cite{Callan:1982ah}. 

\section{Discussion and conclusions}
\label{sec:conclusion}
In this paper, we have performed a detailed calculation for the magnetic and dyonic black holes with electroweak hair  based on the known SM interactions and general relativity. Only spherically symmetric black holes with a magnetic charge $Q=2$ have been considered in this paper. One could extend our calculation to non-spherical larger-$Q$ black holes (for the magnetic monopole states with a large $Q$, see \cite{Ward:1981jb,Houghton:1995bs} for instance). A magnetic RN black hole will evolve into an hMBH if its horizon size $r_H < m_W^{-1}$, which sets an upper bound on the hMBH mass around $9.3\times 10^{35}$~GeV. After subsequent Hawking radiation, it evolves into a (nearly) extremal hMBH. The total mass of an extremal hMBH with $Q=2$ is dominated by its core black hole mass $1.2\times 10^{20}$~GeV, while its hair mass is only $3.6\times 10^{3}$~GeV.  The mass ratio ($\sim 10^{-17}$) of the black hole hair over the total mass is much smaller than the ratio of a single strand of human hair weight over the total body weight $(\sim 10^{-12})$ (this ratio is close to the ratio of a single red-blood cell over the total body weight).  Although the black hole hair weight is small, it extends to a much larger region than the core black hole horizon size. Because of the electroweak symmetry restoration and effective $2d$ massless modes for charged fermions, the $2d$ Hawking radiation is much faster than the ordinary $4d$ case. 

For an extremal hairy dyonic black hole, we have found that its core black hole is just a magnetic one without an electric charge, at least at the leading order in $G v^2$. Although this is different from the ordinary RN black hole case, it can be understood as the large cost of electrostatic energy at the limit of $r \rightarrow r_H$. An integer-charged hDBH can decay very quickly into a lower-charged hDBH and eventually into an hMBH. Therefore, we don't anticipate a stable hDBH existing in the current universe. For a small electric charge $q \ll 2\pi/e^2$, the dyonic hair mass is approximately $q^2\,\times 1.6$~GeV on top of $3.6\times 10^{3}$~GeV for the magnetic hair mass. So, a charge-one hDBH could be produced by cosmic rays hitting a stable hMBH via the processes like 
\beqa
e^- + \mbox{hMBH} &\rightarrow& \mbox{hDBH}^- + \nu_e ~, \\
p + \mbox{hMBH} &\rightarrow& \mbox{hDBH}^+ + n ~,
\eeqa
with the threshold energy of around $1.6$~GeV and the subsequent decays of an hDBH into other cosmic rays. The strength of the signature depends on the populations of hMBHs in the Universe. 

Another interesting avenue beyond our paper is to study the early-universe productions of hairy magnetic black holes. The formation mechanism could be similar to Ref.~\cite{Bai:2019zcd} with additional ingredients to form the monopole configuration based on the SM electroweak sector. As discussed around \eqref{eq:tau-I}, if the initial hMBH is formed as a heavy  non-extremal state with a mass above $\sim 10^{33}$~GeV, its early-stage evaporation time is long enough to leave imprints in the BBN. As also discussed in Ref.~\cite{Bai:2019zcd,Liu:2020vsy,Ghosh:2020tdu,Liu:2020bag}, a binary system with opposite-charged hMBHs could also be  formed during the primordial formation or later structure formation periods. During the merging phase, the hairy dyonic black holes could be generated and its subsequent decays could provide temporal signatures. The final product contains a non-extremal black hole with bright Hawking radiation products that could be searched for by various telescopes. 

It is entertaining to observe that the SM plus general relativity provides us exotic and cosmologically stable states. Along this general direction, one could also study the possibilities of gravitating states without a black hole, like the non-topological soliton stars by Ref.~\cite{Lee:1986tr}, but with the known SM interactions.

\subsubsection*{Acknowledgements}
We would like to thank Joshua Berger and Nicholas Orlofsky for useful discussion. The work is supported by the U.S.~Department of Energy under the contract DE-SC-0017647. The work is in part supported by the Ray MacDonald Intra-mural Award. 

\bibliographystyle{JHEP}
\bibliography{SMBH}

\providecommand{\href}[2]{#2}\begingroup\raggedright\begin{thebibliography}{10}

\bibitem{Misner:1974qy}
C.~W. Misner, K.~Thorne, and J.~Wheeler, {\em {Gravitation}}.
\newblock W. H. Freeman, San Francisco, 1973.

\bibitem{Droz:1991cx}
S.~Droz, M.~Heusler, and N.~Straumann, {\it {New black hole solutions with
  hair}},  {\em Phys. Lett. B} {\bf 268} (1991) 371--376.

\bibitem{Heusler:1991xx}
M.~Heusler, S.~Droz, and N.~Straumann, {\it {Stability analysis of
  selfgravitating skyrmions}},  {\em Phys. Lett. B} {\bf 271} (1991) 61--67.

\bibitem{Bizon:1992gb}
P.~Bizon and T.~Chmaj, {\it {Gravitating skyrmions}},  {\em Phys. Lett. B} {\bf
  297} (1992) 55--62.

\bibitem{Dvali:2016mur}
G.~Dvali and A.~Gu\ss{}mann, {\it {Skyrmion Black Hole Hair: Conservation of
  Baryon Number by Black Holes and Observable Manifestations}},  {\em Nucl.
  Phys. B} {\bf 913} (2016) 1001--1036,
  [\href{http://arxiv.org/abs/1605.00543}{{\tt arXiv:1605.00543}}].

\bibitem{Lee:1991qs}
K.-M. Lee, V.~Nair, and E.~J. Weinberg, {\it {A Classical instability of
  Reissner-Nordstrom solutions and the fate of magnetically charged black
  holes}},  {\em Phys. Rev. Lett.} {\bf 68} (1992) 1100--1103,
  [\href{http://arxiv.org/abs/hep-th/9111045}{{\tt hep-th/9111045}}].

\bibitem{Lee:1991vy}
K.-M. Lee, V.~Nair, and E.~J. Weinberg, {\it {Black holes in magnetic
  monopoles}},  {\em Phys. Rev. D} {\bf 45} (1992) 2751--2761,
  [\href{http://arxiv.org/abs/hep-th/9112008}{{\tt hep-th/9112008}}].

\bibitem{Greene:1992fw}
B.~R. Greene, S.~D. Mathur, and C.~M. O'Neill, {\it {Eluding the no hair
  conjecture: Black holes in spontaneously broken gauge theories}},  {\em Phys.
  Rev. D} {\bf 47} (1993) 2242--2259,
  [\href{http://arxiv.org/abs/hep-th/9211007}{{\tt hep-th/9211007}}].

\bibitem{Maeda:1993ap}
K.-I. Maeda, T.~Tachizawa, T.~Torii, and T.~Maki, {\it {Stability of nonAbelian
  black holes and catastrophe theory}},  {\em Phys. Rev. Lett.} {\bf 72} (1994)
  450--453, [\href{http://arxiv.org/abs/gr-qc/9310015}{{\tt gr-qc/9310015}}].

\bibitem{Lee:1994sk}
K.-M. Lee and E.~J. Weinberg, {\it {Nontopological magnetic monopoles and new
  magnetically charged black holes}},  {\em Phys. Rev. Lett.} {\bf 73} (1994)
  1203--1206, [\href{http://arxiv.org/abs/hep-th/9406021}{{\tt
  hep-th/9406021}}].

\bibitem{Tachizawa:1994wn}
T.~Tachizawa, K.-I. Maeda, and T.~Torii, {\it {NonAbelian black holes and
  catastrophe theory. 2. Charged type}},  {\em Phys. Rev. D} {\bf 51} (1995)
  4054--4066, [\href{http://arxiv.org/abs/gr-qc/9410016}{{\tt gr-qc/9410016}}].

\bibitem{Weinberg:1994eb}
E.~J. Weinberg, {\it {Magnetically charged black holes with hair}},  in {\em
  {13th Symposium on Theoretical Physics: Field Theory and Mathematical
  Physics}}, 6, 1994.
\newblock \href{http://arxiv.org/abs/gr-qc/9503032}{{\tt gr-qc/9503032}}.

\bibitem{Cardoso:2016ryw}
V.~Cardoso and L.~Gualtieri, {\it {Testing the black hole
  \textquoteleft{}no-hair\textquoteright{} hypothesis}},  {\em Class. Quant.
  Grav.} {\bf 33} (2016), no.~17 174001,
  [\href{http://arxiv.org/abs/1607.03133}{{\tt arXiv:1607.03133}}].

\bibitem{Maldacena:2020skw}
J.~Maldacena, {\it {Comments on magnetic black holes}},
  \href{http://arxiv.org/abs/2004.06084}{{\tt arXiv:2004.06084}}.

\bibitem{Chatrchyan:2012ufa}
{\bf CMS} Collaboration, S.~Chatrchyan et~al., {\it {Observation of a New Boson
  at a Mass of 125 GeV with the CMS Experiment at the LHC}},  {\em Phys. Lett.
  B} {\bf 716} (2012) 30--61, [\href{http://arxiv.org/abs/1207.7235}{{\tt
  arXiv:1207.7235}}].

\bibitem{Aad:2012tfa}
{\bf ATLAS} Collaboration, G.~Aad et~al., {\it {Observation of a new particle
  in the search for the Standard Model Higgs boson with the ATLAS detector at
  the LHC}},  {\em Phys. Lett. B} {\bf 716} (2012) 1--29,
  [\href{http://arxiv.org/abs/1207.7214}{{\tt arXiv:1207.7214}}].

\bibitem{Bai:2020spd}
Y.~Bai, J.~Berger, M.~Korwar, and N.~Orlofsky, {\it {Phenomenology of magnetic
  black holes with electroweak-symmetric coronas}},  {\em JHEP} {\bf 10} (2020)
  210, [\href{http://arxiv.org/abs/2007.03703}{{\tt arXiv:2007.03703}}].

\bibitem{Cho:1996qd}
Y.~Cho and D.~Maison, {\it {Monopoles in Weinberg-Salam model}},  {\em Phys.
  Lett. B} {\bf 391} (1997) 360--365,
  [\href{http://arxiv.org/abs/hep-th/9601028}{{\tt hep-th/9601028}}].

\bibitem{Ellis:2016glu}
J.~Ellis, N.~E. Mavromatos, and T.~You, {\it {The Price of an Electroweak
  Monopole}},  {\em Phys. Lett. B} {\bf 756} (2016) 29--35,
  [\href{http://arxiv.org/abs/1602.01745}{{\tt arXiv:1602.01745}}].

\bibitem{tHooft:1974kcl}
G.~'t~Hooft, {\it {Magnetic Monopoles in Unified Gauge Theories}},  {\em Nucl.
  Phys. B} {\bf 79} (1974) 276--284.

\bibitem{Polyakov:1974ek}
A.~M. Polyakov, {\it {Particle Spectrum in the Quantum Field Theory}},  {\em
  JETP Lett.} {\bf 20} (1974) 194--195.

\bibitem{Georgi:1974sy}
H.~Georgi and S.~Glashow, {\it {Unity of All Elementary Particle Forces}},
  {\em Phys. Rev. Lett.} {\bf 32} (1974) 438--441.

\bibitem{Aad:2015zhl}
{\bf ATLAS, CMS} Collaboration, G.~Aad et~al., {\it {Combined Measurement of
  the Higgs Boson Mass in $pp$ Collisions at $\sqrt{s}=7$ and 8 TeV with the
  ATLAS and CMS Experiments}},  {\em Phys. Rev. Lett.} {\bf 114} (2015) 191803,
  [\href{http://arxiv.org/abs/1503.07589}{{\tt arXiv:1503.07589}}].

\bibitem{VanNieuwenhuizen:1975tc}
P.~Van~Nieuwenhuizen, D.~Wilkinson, and M.~Perry, {\it {On a Regular Solution
  of 't Hooft's Magnetic Monopole Model in Curved Space}},  {\em Phys. Rev. D}
  {\bf 13} (1976) 778.

\bibitem{Cho:2016npz}
Y.~Cho, K.~Kimm, and J.~Yoon, {\it {Gravitationally Coupled Electroweak
  Monopole}},  {\em Phys. Lett. B} {\bf 761} (2016) 203--206,
  [\href{http://arxiv.org/abs/1605.08129}{{\tt arXiv:1605.08129}}].

\bibitem{Wald:106274}
R.~M. Wald, {\em {General relativity}}.
\newblock Chicago Univ. Press, Chicago, IL, 1984.

\bibitem{Bai:2019zcd}
Y.~Bai and N.~Orlofsky, {\it {Primordial Extremal Black Holes as Dark Matter}},
   {\em Phys. Rev. D} {\bf 101} (2020), no.~5 055006,
  [\href{http://arxiv.org/abs/1906.04858}{{\tt arXiv:1906.04858}}].

\bibitem{Brihaye:1998cm}
Y.~Brihaye, B.~Hartmann, and J.~Kunz, {\it {Gravitating dyons and dyonic black
  holes}},  {\em Phys. Lett. B} {\bf 441} (1998) 77--82,
  [\href{http://arxiv.org/abs/hep-th/9807169}{{\tt hep-th/9807169}}].

\bibitem{Brihaye:1999nn}
Y.~Brihaye, B.~Hartmann, J.~Kunz, and N.~Tell, {\it {Dyonic nonAbelian black
  holes}},  {\em Phys. Rev. D} {\bf 60} (1999) 104016,
  [\href{http://arxiv.org/abs/hep-th/9904065}{{\tt hep-th/9904065}}].

\bibitem{Julia:1975ff}
B.~Julia and A.~Zee, {\it {Poles with Both Magnetic and Electric Charges in
  Nonabelian Gauge Theory}},  {\em Phys. Rev. D} {\bf 11} (1975) 2227--2232.

\bibitem{Blaer:1981ps}
A.~S. Blaer, N.~H. Christ, and J.-F. Tang, {\it {Anomalous Fermion Production
  by a Julia-Zee Dyon}},  {\em Phys. Rev. Lett.} {\bf 47} (1981) 1364.

\bibitem{Blaer:1981ui}
A.~S. Blaer, N.~H. Christ, and J.-F. Tang, {\it {Fermion Emission From a
  Julia-zee Dyon}},  {\em Phys. Rev. D} {\bf 25} (1982) 2128.

\bibitem{Callan:1982ah}
J.~Callan, Curtis~G., {\it {Disappearing Dyons}},  {\em Phys. Rev. D} {\bf 25}
  (1982) 2141.

\bibitem{Sonoda:1983qh}
H.~Sonoda, {\it {Decay of a Dyon}},  {\em Nucl. Phys. B} {\bf 238} (1984)
  259--276.

\bibitem{Witten:1979ey}
E.~Witten, {\it {Dyons of Charge e theta/2 pi}},  {\em Phys. Lett. B} {\bf 86}
  (1979) 283--287.

\bibitem{Ward:1981jb}
R.~Ward, {\it {A Yang-Mills Higgs Monopole of Charge 2}},  {\em Commun. Math.
  Phys.} {\bf 79} (1981) 317--325.

\bibitem{Houghton:1995bs}
C.~J. Houghton and P.~M. Sutcliffe, {\it {Tetrahedral and cubic monopoles}},
  {\em Commun. Math. Phys.} {\bf 180} (1996) 343--362,
  [\href{http://arxiv.org/abs/hep-th/9601146}{{\tt hep-th/9601146}}].

\bibitem{Liu:2020vsy}
L.~Liu, O.~Christiansen, Z.-K. Guo, R.-G. Cai, and S.~P. Kim, {\it
  {Gravitational and electromagnetic radiation from binary black holes with
  electric and magnetic charges: Circular orbits on a cone}},  {\em Phys. Rev.
  D} {\bf 102} (2020), no.~10 103520,
  [\href{http://arxiv.org/abs/2008.02326}{{\tt arXiv:2008.02326}}].

\bibitem{Ghosh:2020tdu}
D.~Ghosh, A.~Thalapillil, and F.~Ullah, {\it {Astrophysical hints for magnetic
  black holes}},  \href{http://arxiv.org/abs/2009.03363}{{\tt
  arXiv:2009.03363}}.

\bibitem{Liu:2020bag}
L.~Liu, O.~Christiansen, Z.-K. Guo, R.-G. Cai, and S.~P. Kim, {\it
  {Gravitational and electromagnetic radiation from binary black holes with
  electric and magnetic charges: Elliptical orbits on a cone}},
  \href{http://arxiv.org/abs/2011.13586}{{\tt arXiv:2011.13586}}.

\bibitem{Lee:1986tr}
T.~Lee and Y.~Pang, {\it {Fermion Soliton Stars and Black Holes}},  {\em Phys.
  Rev. D} {\bf 35} (1987) 3678.

\end{thebibliography}\endgroup
\end{document}